\documentclass[12pt]{iopart}%

\usepackage{amsfonts}%
\usepackage{amssymb}%
\usepackage{graphicx}
\newtheorem{theorem}{Theorem}
\newtheorem{acknowledgement}[theorem]{Acknowledgement}

\begin{document}

\title{Orthonormal Filters for Identification in Active Control Systems}
\author{Dirk Mayer \footnote{present address: Fraunhofer LBF, Darmstadt, Germany}}
\ead{dirk.mayer@lbf.fraunhofer.de}
\

\begin{abstract}
Many active noise and vibration control systems require models of the control paths. When the controlled system changes slightly over time, adaptive digital filters for the identification of the models are useful. This paper aims at the investigation of a special class of adaptive digital filters: Orthonormal filter banks possess the robust and simple adaptation of the widely applied Finite Impulse Response (FIR) filters, but at a lower model order, which is important when considering implementation on embedded systems. However, the filter banks require prior knowledge about the resonance frequencies and damping of the structure. This knowledge can be supposed to be of limited precision, since in many practical systems, uncertainties in the structural parameters exist. In this work, a procedure using a number of training systems to find the fixed parameters for the filter banks is applied. The effect of uncertainties in the prior knowledge on the model error is examined both with a basic example and in an experiment. Furthermore, the possibilities to compensate for the imprecise prior knowledge by a higher filter order are investigated. Also comparisons with FIR filters are implemented in order to assess the possible advantages of the orthonormal filter banks. Numerical and experimental investigations show that significantly lower computational effort can be reached by the filter banks under certain conditions.
\end{abstract}

\submitto{\SMS}
\maketitle

\section{Introduction}
Active control of noise and vibration has been applied to numerous applications in the last decades. Especially when periodic disturbances from engines or other machinery have to be canceled, the Filtered-X-Least Mean Square (FXLMS) algorithm has been found performing well \cite{widrow_1985} \cite{kuo_1996} \cite{elliott_2000}. This special class of adaptive filter systems requires input-output models of the control paths, i.e. the transfer function matrix from the actuators to the sensors.\\
Deviations of the model from the actual control path characteristics limit stability and performance of the FXLMS control system. Thus, regular update of the models is desirable in order to maintain a sufficient performance and stability of the control system. A possible solution is the identification of secondary path models for different operational conditions or other parameters and apply those later in the control system with respect to the actual condition \cite{Zhao.2015}. Also, on-line modeling of the secondary path has been considered \cite{kuo_1997}\cite{hesselbach_2009}. In \cite{pu_new_2014} and \cite{mingzhang_2001} a review of implementations and proposals for further methods can be found. In all cited works, the identification of the secondary path is implemented with adaptive digital filters. Due to their simple structure and good convergence properties, finite impulse response (FIR) filters are preferred in most of the documented applications. Since FIR filter coefficients represent the samples of the impulse response, high filter orders for control path modeling are usually required in order to gain sufficient precision. This problem rises as well in resonant mechanical structures and acoustic paths \cite{liu_2012} \cite{rohlfing_2014}. 
Alternatively, infinite impulse response (IIR) filters can be applied. They incorporate internal feedback, which allows for modeling of resonant systems with a low number of filter parameters. Adaptation of the system poles can raise some issues, though. With the output error scheme, which aims at directly minimizing the error between the adaptive filter output and the output of the system to be identified, the adaptation algorithm might get stuck in local minima, i.e. the solution depends on the proper choice of start values. 
Alternatively, an equation error can be formulated. This includes a filtering of the system output with an FIR filter constructed from the denominator coefficients of the model; this way the adaptive IIR filtering problem is deduced to the adaptation of two FIR filters \cite{elliott_2000}. This configuration usually converges better, but might procure unstable models which cannot be used in a control system \cite{widrow_1985}. Furthermore, in the presence of measurement noise, enhanced filter structures have to used in order to prevent bias in the estimated parameters \cite{lopez-valcarce_algorithm_2003}. In comparative studies, IIR filters are found to perform better than FIR filters, however, problems with stability and convergence are documented \cite{kegerise_2002}.\\
In order to combine the good convergence of the transversal filter with the low required number of coefficients, a third type of adaptive filters has been proposed for on-line system identification, which uses a bank of fixed-pole IIR filters. Adaptation is applied only to the weights at the output signals of those filters, i.e. a linear-in-parameter model structure is used, which provides similar convergence properties like adaptive FIR filters. 
A quite straightforward approach to the design of such filter banks is the fixed-pole adaptive filter \cite{williamson_globally_1996}. However, most filter banks are designed using series of orthonormal basis functions \cite{vandenHof_1994}. Due to the orthonormal structure of the resulting filter bank, a faster convergence during adaptation is provided compared to other structures \cite{ninness_practical_2000}. Adaptation of the filter coefficients can be implemented with straightforward stochastic Least Mean Square algorithms \cite{mayer_application_2001}. An interesting feature is that those orthonormal filter structures have been observed to be more robust against noise in the identification process than IIR models \cite{Lemma_2010}. For the identification of flexible mechanical structures, Kautz functions \cite{kautz_transient_1954} are well suited, since they represent exponentially decaying oscillations which resemble the impulse response of a mechanical structure. They have been applied to replace FIR filters in order to reduce memory usage in terms of delay elements due to the significantly lower filter order needed \cite{friedman_approximating_1981} \cite{wahlberg_identification_1991}, and also to the identification of flexible structures \cite{nalbantoglu_system_2003},\cite{da_silva_non-parametric_2011},\cite{mayer_application_2001}.\\
The main challenge in the application of the orthonormal filter banks reamins the choice of the fixed pole locations by incorporation of a priori knowledge. If this matches the actual poles well, a Kautz filter of low order will provide a good model. Experimental system identification procedures can be applied to gather the necessary knowledge of the system poles \cite{nalbantoglu_system_2003}. Optimality conditions for the poles of the Kautz filter were derived from conditions on the filter weights in case of a Kautz filter with a single pole \cite{den_brinker_optimality_1996}. Alternatively, in \cite{da_silva_non-parametric_2011} the application of numerical optimization algorithms to this task has been proposed. Also, adaptation schemes for the poles have been developed, but at the price of higher computational efforts for the filter system \cite{friedman_approximating_1981}.\\
In most of those works it assumed that the necessary a priori knowledge can be gained from data taken at the system to be identified. However, due to the uncertainties mentioned above, in an active control system most likely some parameters will change during operation or from specimen to specimen which the system is applied to, and the a priori knowledge will not perfectly match the current system poles. In several works a priori knowledge was gained from several \textsl{example systems} of a certain type and with similar characteristics. After that, the adaptive filter bank was used to identify other realizations of the systems. In \cite{williamson_globally_1996}\cite{zimmermann_performance_1993}, a set of slightly different butterworth bandpass filters is used as a case study. Another study included measured impulse responses of digital communication channels with similar characteristics as example systems \cite{kaelin_simplified_1995}.\\
In the work presented here, the application of Kautz filters to the identification of mechanical structures is studied with a focus on uncertainty in the a priori knowledge of the system's poles. The idea to use the properties of several example systems is taken on and transferred to the identification of mechanical systems.
In the following, the system identification procedure with adaptive Kautz filters is briefly summarized. First investigations of the performance are conducted with a single degree-of-freedom (SDOF) system, also regarding the situation that the apriori knowledge does not perfectly match the actual system poles. It can be expected, that in this case a higher order Kautz filter will be needed; thus comparisons of the performance to the well established FIR filters are implemented.
\section{Kautz filter banks}
\begin{figure}
\includegraphics[width=0.7\textwidth]{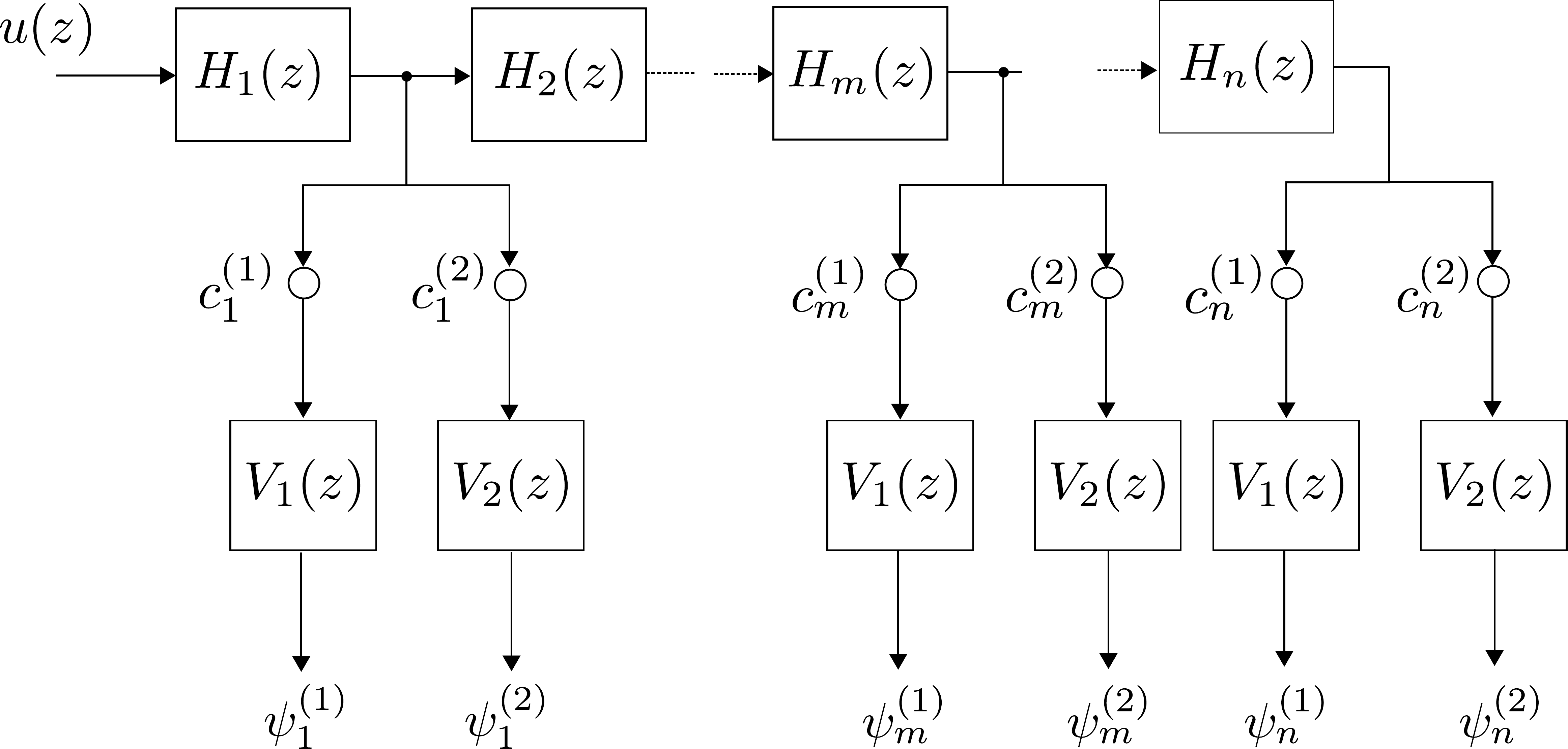}
\caption{Realization of the Kautz filter bank}
\label{kautz_filterbank}
\end{figure}

\begin{figure}
\includegraphics[width=0.7\textwidth]{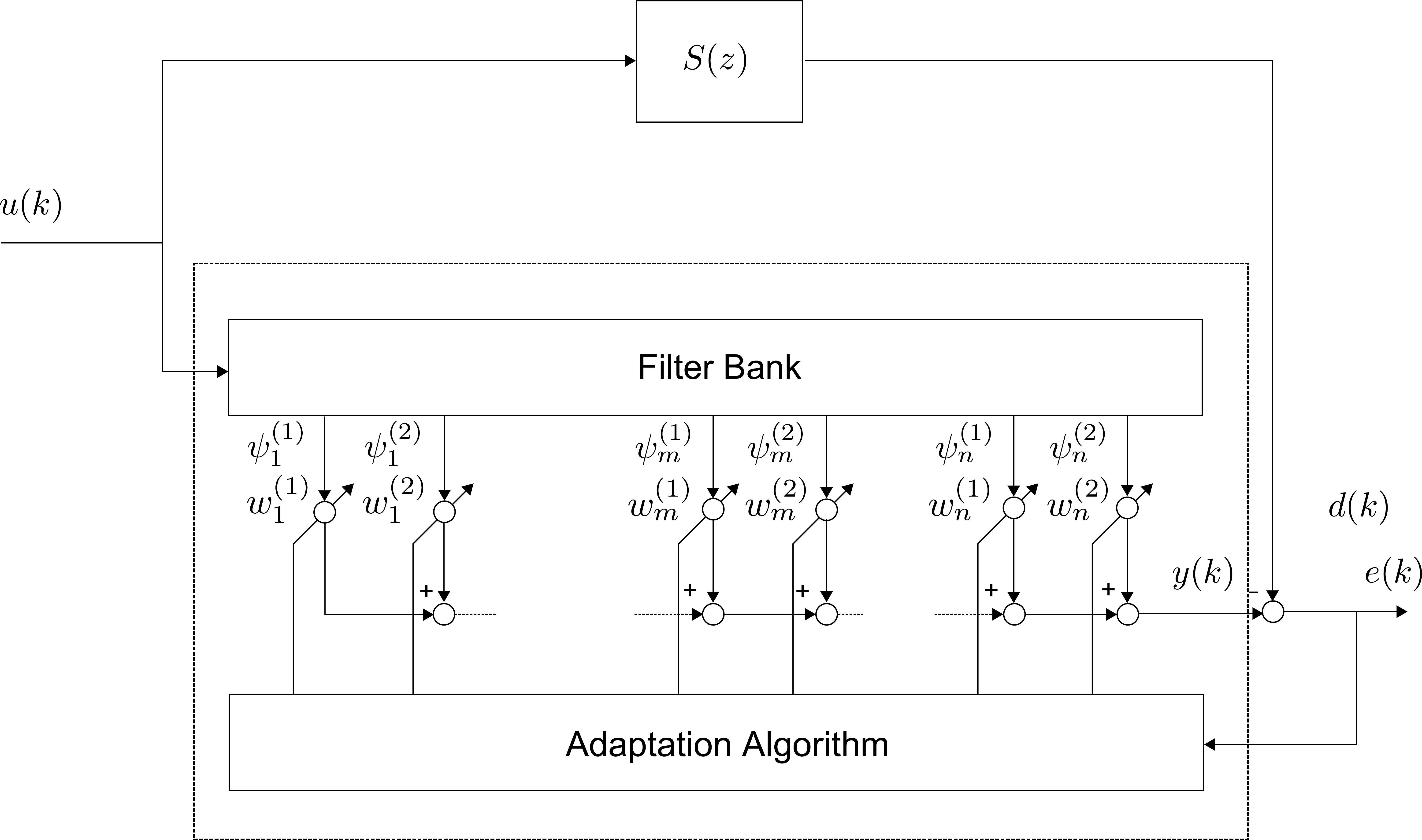}
\caption{System identification with an adaptive filter bank}
\label{alc_sysid}
\end{figure}

The basic idea behind orthonormal filter banks is the approximation of a given impulse response $h(n)$ by a finite series of orthonormal functions. As stated above, in the considered case of modeling flexible structures, the Kautz functions are supposed to work well. Since usually several resonances are expected, Kautz functions with multiple, conjugated complex pole pairs are used. Thus, the Kautz series considered here comprises an even number of functions $\psi_n^{(1,2)}$ and coefficients $w_n^{(1,2)}$. The approximation reads:
\begin{equation}
	h(k) \approx \hat{h}(k) = \sum_{n=1}^N{w_n^{(1)} \psi_n^{(1)}(k) + w_n^{(2)} \psi_n^{(2)}(k)}
\end{equation}
The functions $\psi_n^{(1,2)}$ are defined as the inverse z-transform of the following transfer functions $\Psi_n^{(1,2)}$ \cite{paatero_kautz_2003}:
\begin{equation}
	\Psi_n^{(1)} (z) = c_n^{(1)} \frac{z (z-1)}{(z-p_n)(z-p_n^*)} \prod_{m=1}^n{\frac{(1-p_m z)(1-p_m^* z)}{(z-p_m)(z-p_m^*)}}
\end{equation}
\begin{equation}
	\Psi_n^{(2)} (z) = c_n^{(2)} \frac{z (z+1)}{(z-p_n)(z-p_n^*)} \prod_{n=1}^n{\frac{(1-p_m z)(1-p_m^* z)}{(z-p_m)(z-p_m^*)}}
\end{equation}
Here, $p_m$ is a set of poles inside the unit circle constructed from the a priori knowledge. Procedures for choosing this set are described in the next section. 
 The constants $c_n^{(1,2)}$ are necessary for proper scaling of the functions:
\begin{equation}
	c_n^{(1,2)} = \sqrt{\frac{(1+p_n^2)(1 \pm p_n^*)(1-\left|p_n\right|^2 )}{2}}
\end{equation}
Thus, the Kautz functions form an orthonormal set of base functions, fulfilling the condition:
\begin{equation}
(\psi_l^{(i)},\psi_m^{(j)})=\sum_{k=0}^{\infty} \psi_l^{(i)}(k) \psi_m^{(j)}(k) = \left\{
\begin{array}{ll}
1 & \mbox{if $l=m, i=j$}; \\
0 & \mbox{else.}	
\end{array}
\right.
\end{equation}
The Kautz functions can be implemented with a filter bank as shown in Figure \ref{kautz_filterbank}, where
\begin{equation}
H_1(z) = \frac{z^2}{(z-p_1)(z-p_1^*)}
\end{equation}
\begin{equation}
H_m(z) = \frac{(1-p_{m-1}z)(1-p_{m-1}^* z)}{(z-p_m)(z-p_m^*)}
\end{equation}
\begin{equation}
  V_1(z) = \frac{1-z}{z} 
\end{equation}
\begin{equation}
 V_2(z) = \frac{1+z}{z}
\end{equation}
To implement the adaptive Kautz filter, the outputs of the filter bank are used as inputs to an adaptive linear combiner \cite{widrow_1985}. The system identification with this arrangement is shown in Figure \ref{alc_sysid}. The output $y(k)$ of the Kautz filter is a linear combination of the output signals due to a test signal $u(k)$ with the filter weights $ w_n^{(1,2)}$
\begin{equation}
	y(k) = \sum_{n=1}^N{w_n^{(1)} \psi_n^{(1)}(k) + w_n^{(2)} \psi_n^{(2)}(k)}
\end{equation}
The desired signal $d(k)$ is the output of the system $S(z)$ to be identified due to the same test signal $u(k)$. The difference of both output signals gives the error signal $e(k)$:
\begin{equation}
e(k)  = d(k) -y(k)
\end{equation}
The goal of the adaptation is to minimize the power of this error signal. As shown in \cite{widrow_1985}, the straightforward stochastic adaptive LMS algorithm can be applied:
\begin{equation}
	w_n(k+1)^{(1,2)} = w_n(k)^{(1,2)} + \mu \psi_n^{(1,2)}(k) e(k)
\end{equation}
Due to the orthonormality of the inputs to the adaptive linear combiner, the adaptation of each filter weight is independent from the others. This enables a high adaptation rate of the adaptive filter. 
\section{Choice of poles}
\begin{figure}
\includegraphics[width=0.5\textwidth]{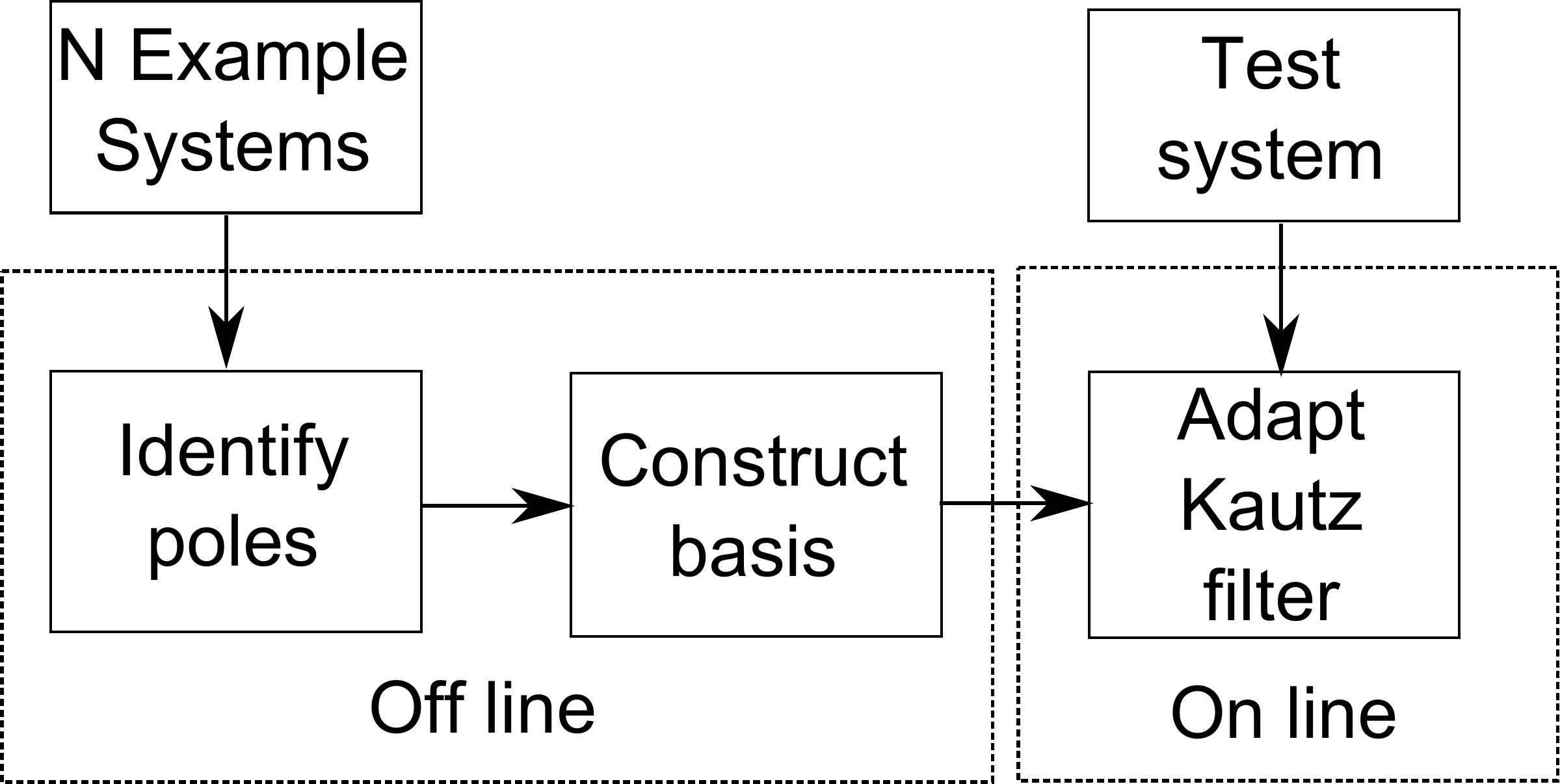}
\caption{Procedure for system identification with adaptive filter banks}
\label{kautz_procedure}
\end{figure}

A low approximation error with a short Kautz filter requires the incorporation of knowledge on the system poles $p_m$ into the Kautz functions $\Psi_n^{(1,2)} (z)$. Similar to \cite{nalbantoglu_system_2003}, it is assumed that several example systems are available for experimental analysis to estimate the system poles. For instance, several specimen of a mechanical structure which are slightly different due to tolerances in manufacturing processes or materials, or of one structure under several operating conditions like temperatures could be considered. 
At the moment, it should be assumed that a set of representative training systems is available. For artificially created sets of training systems, the statistic properties are well known \cite{davidson_1991} \cite{zimmermann_performance_1993}. In a real situation an investigation of the uncertainties in the parameters in order to get an estimate of the existing parameter variations and find a proper set of training systems might be necessary. However, usually a limited number of training systems is available and completely used for the procedure of finding proper poles of orthonormal filters \cite{kaelin_simplified_1995}. 
The orthonormal basis is chosen based on the results of the analysis of the training systems. Finally, the filter bank is deployed to the on-line identification of the test systems by adaptation of the filter weights. The procedure is illustrated in Figure \ref{kautz_procedure}. 
%
%
A practical way to examine the training systems is the identification of the impulse response, e.g. with an adaptive FIR filter of sufficient order. The impulse responses are used to construct a Single input Multiple Output (SIMO) system, which then includes the variations between the training systems in its different channels. A modified Prony method is applied to identify the poles of this SIMO system \cite{williamson_globally_1996}. The idea of this strategy is to include the information on the variation between the training sets into a single system and find a proper set of poles which represents those characteristics. Of course, if one or several training sets are rather poorly representing the expected variation, e.g. because data is taken in very uncommon conditions, the variations are due to damages and not tolerances etc., this method might fail to procure a useful set of poles. A detailed investigation on this issue, however, is beyond the scope of this work.
\\
To this end, the $M$ impulse responses, each of length $l$, are arranged in a matrix. Presuming a model order $N$, the following equation can be solved for the vector $[\alpha_1 \dots \alpha_N]$:
\begin{equation}
	\left[ 
	\begin{array}{lll}
		h_1(N-2) & \cdots & h_1(0) \\
		\vdots &          & \vdots \\
		h_1(l-1) & \cdots & h_1(l-N) \\
		h_2(N-2) & \cdots & h_2(0) \\
		\vdots &          & \vdots \\
		h_2(l-1) & \cdots & h_2(l-N) \\
		\vdots &          & \vdots \\
		h_M(N-2) & \cdots & h_M(0) \\
		\vdots &          & \vdots \\
		h_M(l-1) & \cdots & h_M(l-N) 
		\end{array}
		\right]
		\left[ 
		\begin{array}{c}
		\alpha_1\\
		\alpha_2\\
		\vdots\\
		\alpha_N
		\end{array}
		\right] = 
			\left[ 
		\begin{array}{l}
    h_1(N-1) \\
		\vdots \\
		h_1(l) \\
		h_2(N-1) \\
		\vdots \\
		h_2(l) \\
		\vdots \\
		h_M(N-1) \\
		\vdots\\
		h_M(l) 
		\end{array}
		\right] 		
\end{equation}
Similar to the single channel Prony method, the poles $z_n$ are found by solving:
\begin{equation}
	z^N+\alpha_1 z^{N-1}+\alpha_2 z^{N-2}+\ldots+\alpha_{N-1}z+\alpha_N = 0
\end{equation}
In addition to the knowledge gathered from the Prony algorithm, also general knowledge about the system to be identified can be helpful:
In real systems, a certain noise level is present in the identified impulse responses. In this case, the Prony method procures also poles that are physically not meaningful. In the experiments described later on, poles with $|z|<1$ were skipped when constructing the orthonormal basis for the filter bank; also real-valued poles were removed, because in an elastic mechanical structure with low damping no overdamped resonances are expected. In those cases, the Prony algorithm is started again with an increased order until $N$ suitable poles are available. The estimated poles are used to construct the orthonormal basis for the filter bank. This bank is deployed to identify the test system by adaptation of the weights $w_n$.
It should be noted that also other strategies for the definition of the poles of the Kautz filter bank may be applied, e.g. using data from experimental modal analysis.\\
%
%
\section{Comparison with other filter structures}
Obviously, the realization of the Kautz filter bank is computationally more complex than a FIR filter with respect to the filter length, since simple delays are replaced by second order filters. As shown in Table \ref{table_efforts}, the effort for the implementation of a Kautz filter bank is roughly three times higher compared to a FIR filter, which has to be considered when implementing Kautz filters of higher order.
\begin{table}
	\centering
		\begin{tabular}{|c||c|c|c|c|}
		\hline
			Filter type  	& Additions & Multiplications & Divisions & Storage (Memory) \\
			\hline
			\hline
			FIR 					& $n+1$       &   $ n $       & 0 & $n$ \\
			\hline
			Kautz         & $5+3(n-3)$ & $3.5 (n-3) +8$ & 0 & $ 3n $\\
			\hline		
			
		\end{tabular}
		\caption{Efforts for the implementation of the FIR and Kautz filters for a filter of order $n$ }
		\label{table_efforts}
\end{table}
\section{Basic investigations}
\begin{figure}
\begin{tabular}{cc}
\includegraphics[width=0.5\textwidth]{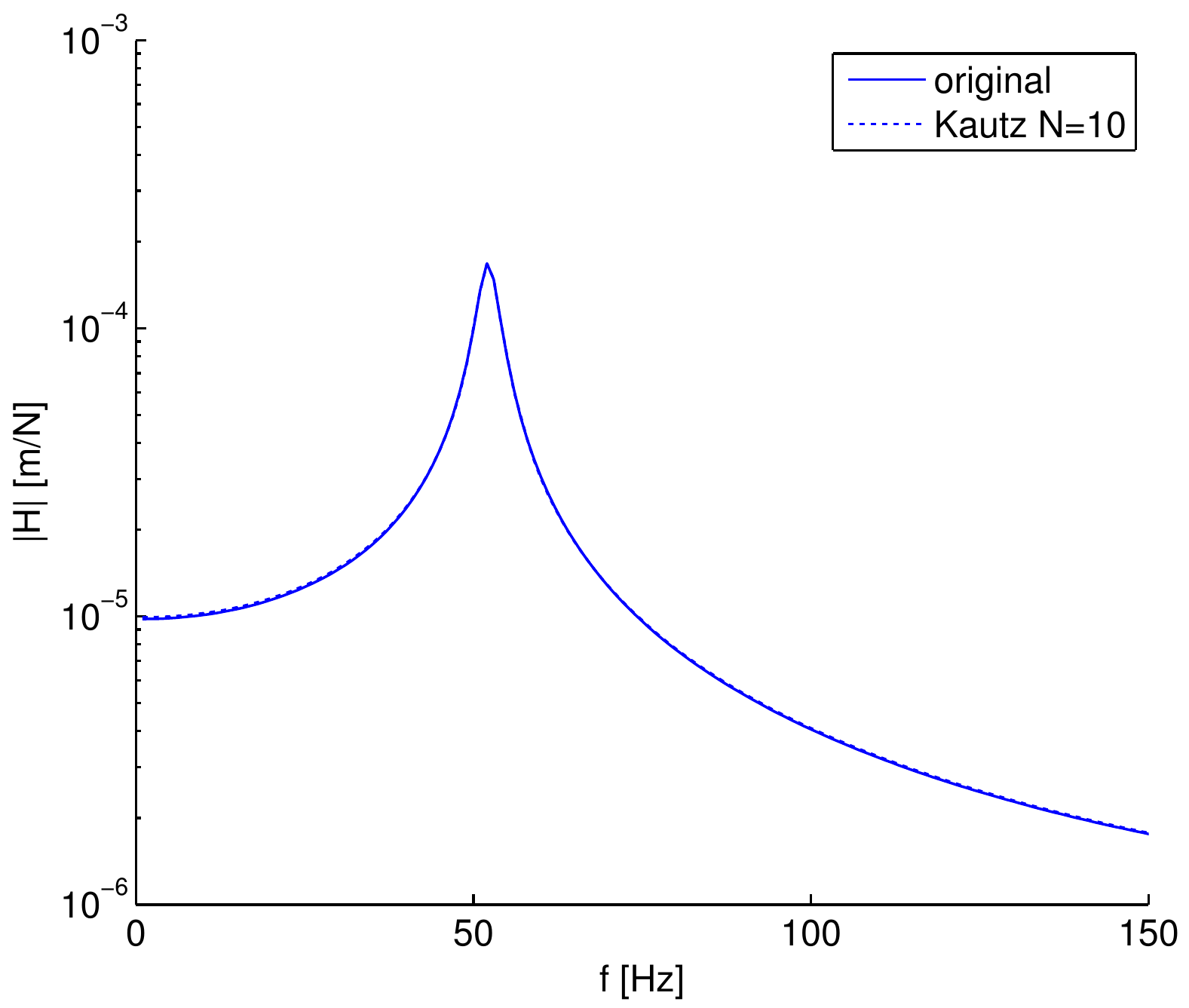}&\includegraphics[width=0.5\textwidth]{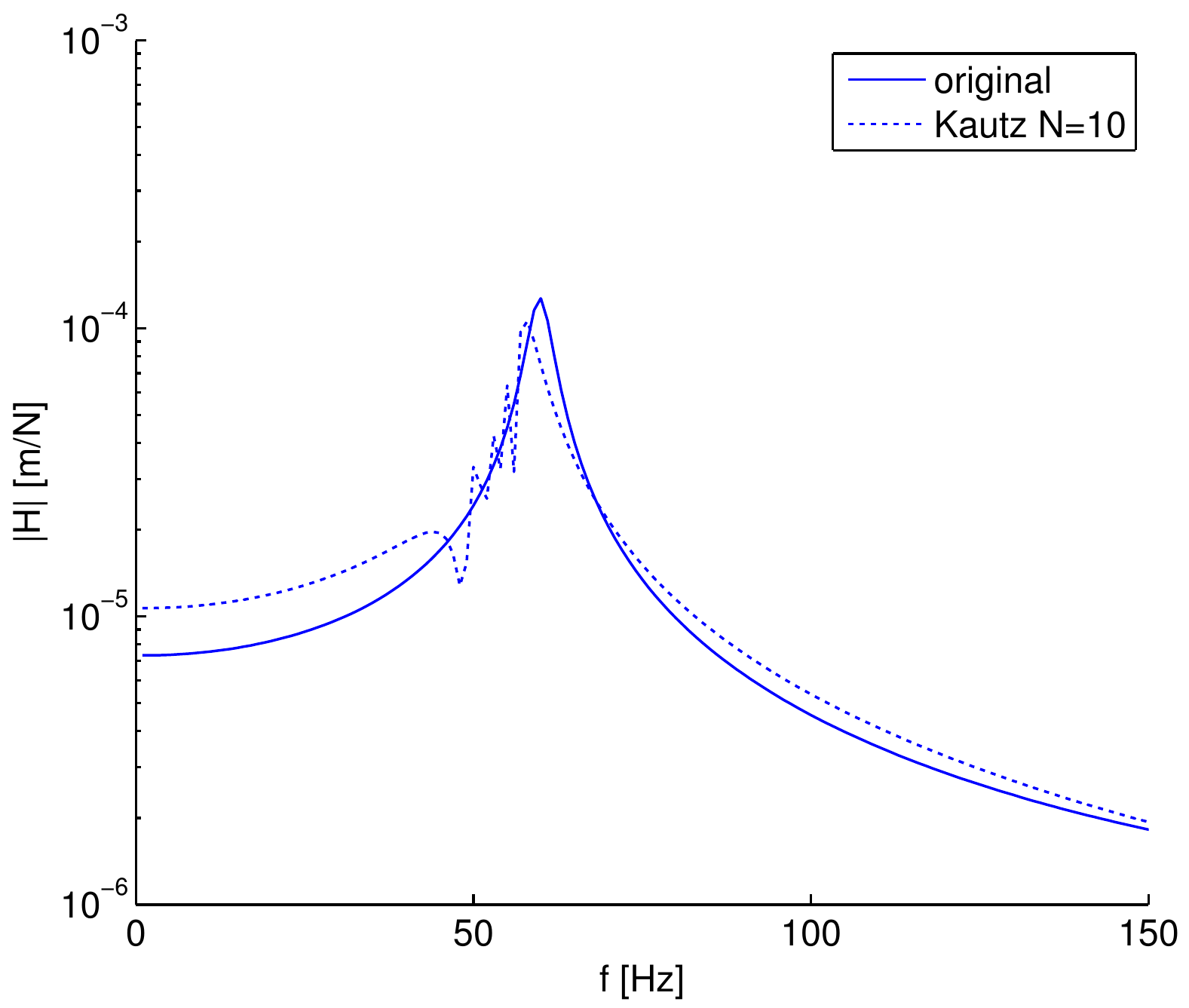}
\end{tabular}
\caption{Approximation of the test system with a Kautz filter of order 10, best case (left) and worst case (right)}
\label{frf_best}
\end{figure}

\begin{figure}
\begin{tabular}{cc}
\includegraphics[width=0.5\textwidth]{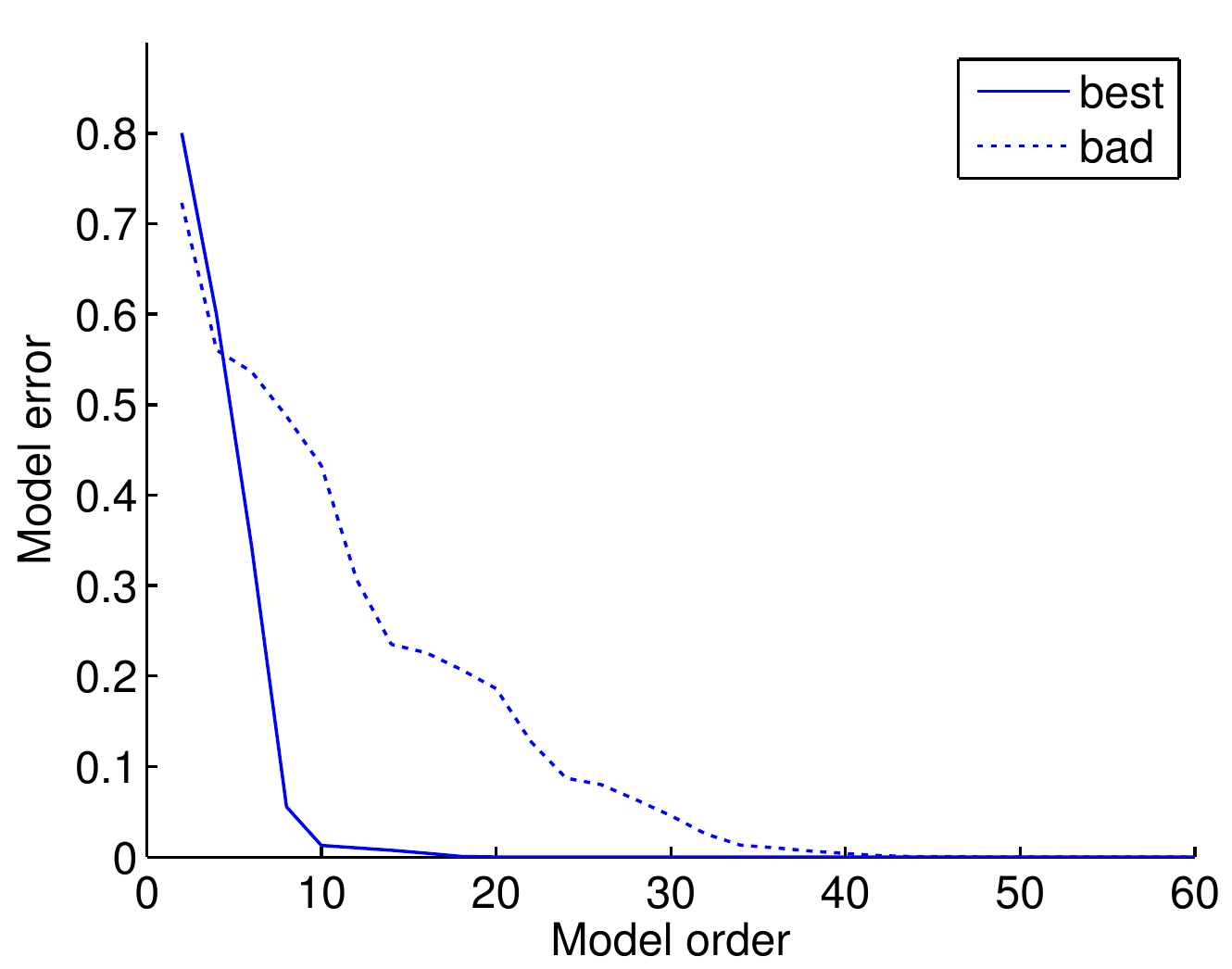}&\includegraphics[width=0.5\textwidth]{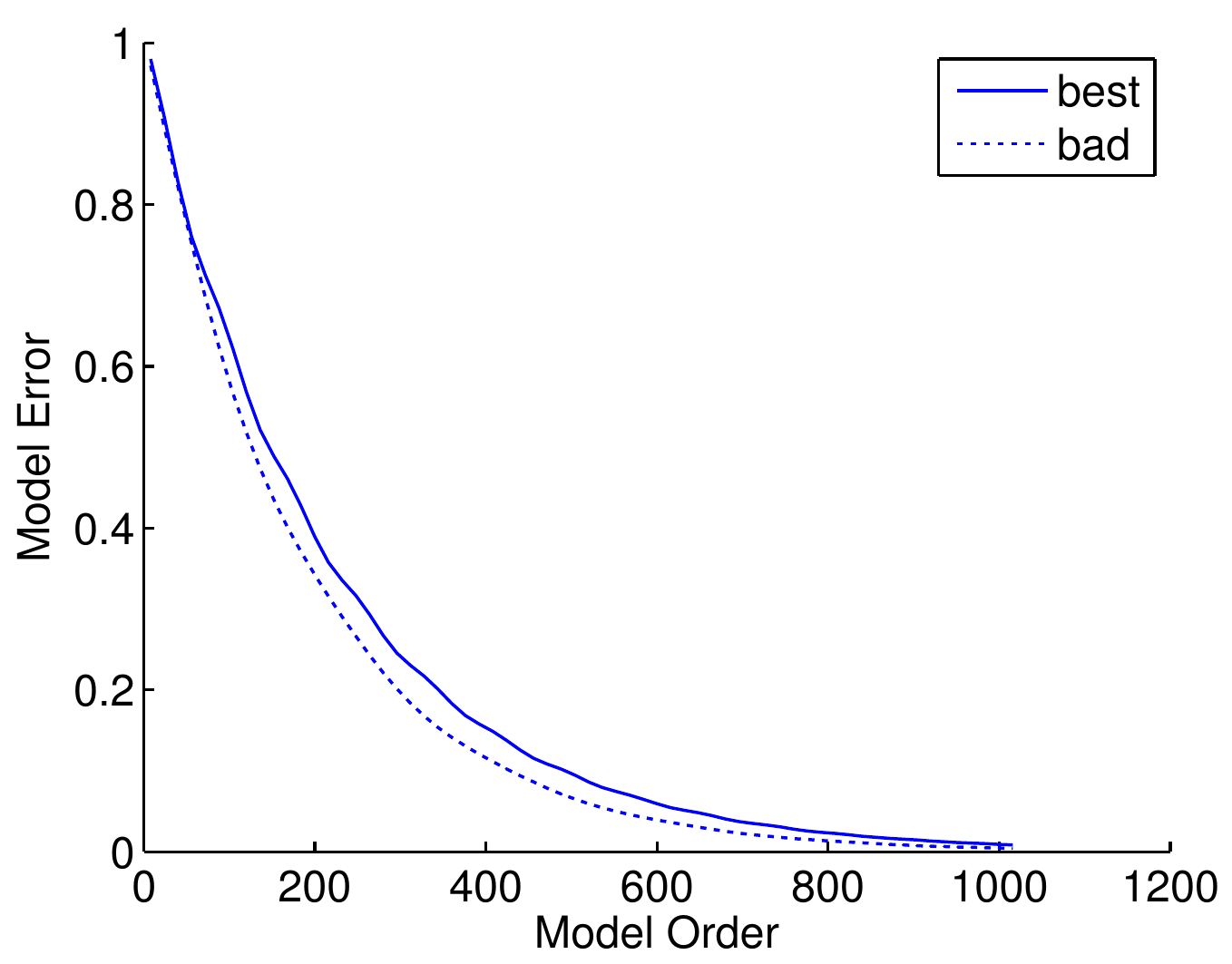}
\end{tabular}
\caption{Approximation of the test system with Kautz filters of ascending orders (left) and FIR filters (right)}
\label{app_error}
\end{figure}

In this section, the properties of the Kautz filter bank are studied by a basic example. The goal is to assess the performance of the filter bank when identifying flexible systems. The characteristics like resonance and damping should be known in advance, but with some degree of uncertainty.\\
To this end, a single-degree-of-freedom (SDOF) oscillator is considered, characterized by its mass, stiffness and damping ratio. In the following, uncertainties in the resonance frequency and damping are considered.\\
The continuous time representation of a SDOF system with a force input and displacement output signal reads:
\begin{equation}
	\frac{x(j\omega)}{F(j\omega)}=\frac{1/m}{-\omega^2 + 2j\omega_0 \theta_0 \omega + \omega_0^2},
\end{equation}
where $m$ represents the mass, $\omega_0 = \sqrt{k/m}$ the resonance frequency and $\theta_0$ the damping coefficient. For system identification with digital filter structures, this system is shifted to the $z$-domain by the impulse invariance transformation. With the sampling interval $t_s$, the poles of this system in the $z$-plane are:
\begin{equation}
	 z_{1,2} = e^{(-\theta_0+j)\omega_0\sqrt{1-\theta_0^2} t_s}.
\end{equation}
The uncertainties in the parameters should be represented by a normal distribution with the mean values $\omega_0$, $\theta_0$ and the standard deviations $\Delta_\omega$, $\Delta_\theta$. Thus, the probability density function for resonance frequency and damping reads:
\begin{equation}
	P_\omega = \omega_0 + \frac{1}{\sqrt{2\pi\Delta_\omega^2}}e^{-\frac{(\omega - \omega_0)^2}{2\Delta_\omega^2}}
\end{equation}
\begin{equation}
  P_\theta = \theta_0 + \frac{1}{\sqrt{2\pi\Delta_\theta^2}}e^{-\frac{(\theta - \theta_0)^2}{2\Delta_\theta^2}}
\end{equation}
As an example, a single-degree-of-freedom system with a nominal resonance frequency of 50Hz and a damping coefficient of $0.03$ is chosen. A standard deviation of $0.05$ is assumed for the resonance and damping. 
First a set of five example or training systems is generated with the defined random distribution. This set is used for the definition of the Kautz filter poles. After setting the fixed poles of the Kautz filter bank, a set of test systems is generated. This set is used to evaluate the performance of the adaptive filters. Two test systems are studied. First, a test system which exactly matches the parameters of the SDOF-system, which is approximately the best case. Second, a system with a resonance frequency of with 15\% deviation to the mean value. For a relative standard deviation (or coefficient of variation) of 0.05, over 99\% of the systems can be expected to have a deviation in this range. This may be considered a worst case in a practical situation (referred to as \textit{bad case} in the following): Considering the SDOF system, the stiffness depends to the square of the resonance frequency. A deviation of $\pm 15\%$ thus corresponds to a deviation of the stiffness in a range from -27.8\% to +32.2\%. Compared to deviations observed at real automotive structures \cite{durand_2008} or composite parts \cite{lekou_2008}, this seems a realistic assumption.
The parameters for the training and test systems are summarized in Table \ref{num_example}.\\
\begin{table}
	\centering
		\begin{tabular}{|c||c|c|c|c|}
		  \hline
			Training Systems & $f_0$ [Hz] & $\theta$ [-] \\
			\hline
			               
	& 53.5 & 0.031 \\
  & 50.7 & 0.031 \\
  & 50.5 & 0.030 \\
  & 54.0 & 0.030 \\
  & 48.0 & 0.028 \\
	  \hline
    \hline
 Test Systems & & \\
		\hline
 Best case & 50.0 & 0.030 \\
 Bad case  & 57.5 & 0.030 \\
		\hline    
		\end{tabular}
		\caption{Parameters for the numerical simulation of the SDOF system identification with Kautz filters}
		\label{num_example}
\end{table}
Adaptive Kautz filters are approximated in a numerical simulation using a sampling frequency of 500Hz.
The approximation error is defined to:
\begin{equation}
e = \frac{\sum_{k=1}^K{({h}(k)-\hat{h}(k))^2}}{\sum_{k=1}^K({h}(k))^2}, 
\end{equation}
where $K$ denotes the maximum length of the impulse response considered; for the example, a length of 500 samples was used.
Figure \ref{frf_best} shows an example for the approximation with a model of order 10, both for the considered best and worst case. The respective approximation errors are 0.01 and 0.43, respectively. Obviously, the quality of the a priori knowledge has a strong influence on the estimation error. 
The described simulations are repeated with 1000 sets of 5 randomly chosen training systems with the same statistical properties as above. Again, models of increasing order are approximated in numerical simulations for both cases.  
Since the identification does not procure more poles than existing in the training system, just 5 distinguished poles can be used for defining the poles of the Kautz filter bank. For Kautz filters of a higher order, this set of poles is utilized repeatedly in series connection.\\
Figure \ref{app_error} shows the mean approximation error for both the best and the worst case, each for models of increasing order. Both Kautz models and FIR models were approximated. As expected, the FIR filters have a far higher order than the Kautz filters. Even when considering the higher effort for the implementation of the Kautz filter, savings in computation effort can be expected. However, also the disadvantage of Kautz filters can be noticed from Figure \ref{app_error}: The approximation error for a given order is strongly depending to the deviation of the test system to the training system set: In the bad case, the model order has to be chosen about two times higher to reach the same error.
\section{Application example}
The described procedure is applied to a more complex example using measured data from a laboratory experiment. The test set up comprises a steel truss structure which was utilized for other experiments on active and passive vibration control before \cite{mayer_development_2012}. An electrodynamic shaker with integrated impedance head (Wilcoxon F3/Z602WA) is used for excitation and measurement, an Ono Sokki FFT analyzer for data acquisition and calculation of the frequency response functions. 
The data is acquired with a sampling rate of 500Hz, and the calculated FRFs have 3184 lines. After wards, the impulse response functions are calculated by first removing the low-frequency components below 10 Hz, which are not properly excited by the shaker, and an inverse FFT.


\begin{figure}
	\centering
		\includegraphics[width=0.50\textwidth]{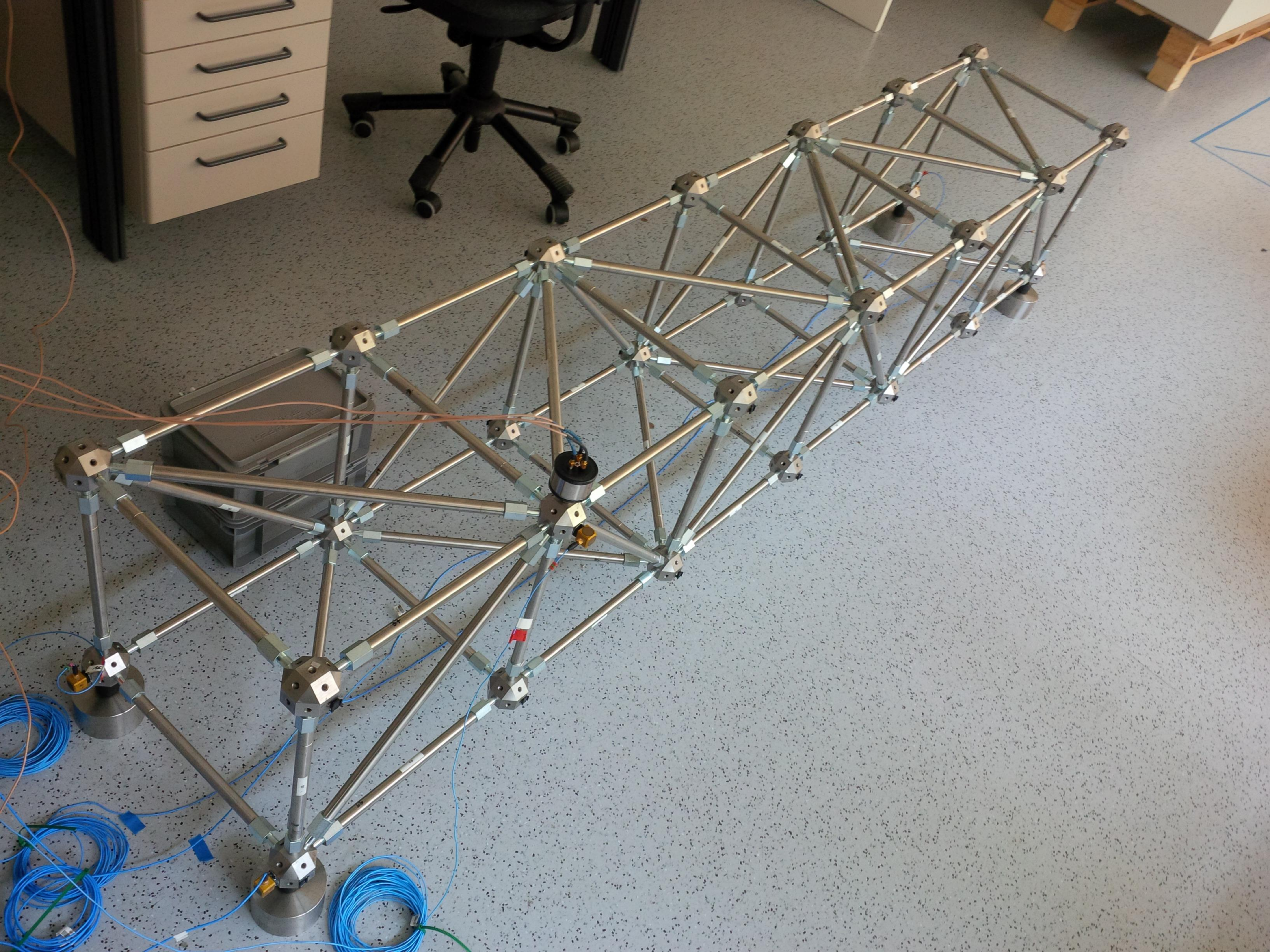}
	\caption{Experimental set up}
	\label{fig:truss_setup}
\end{figure}
\begin{figure}
	\centering
		\includegraphics[width=0.50\textwidth]{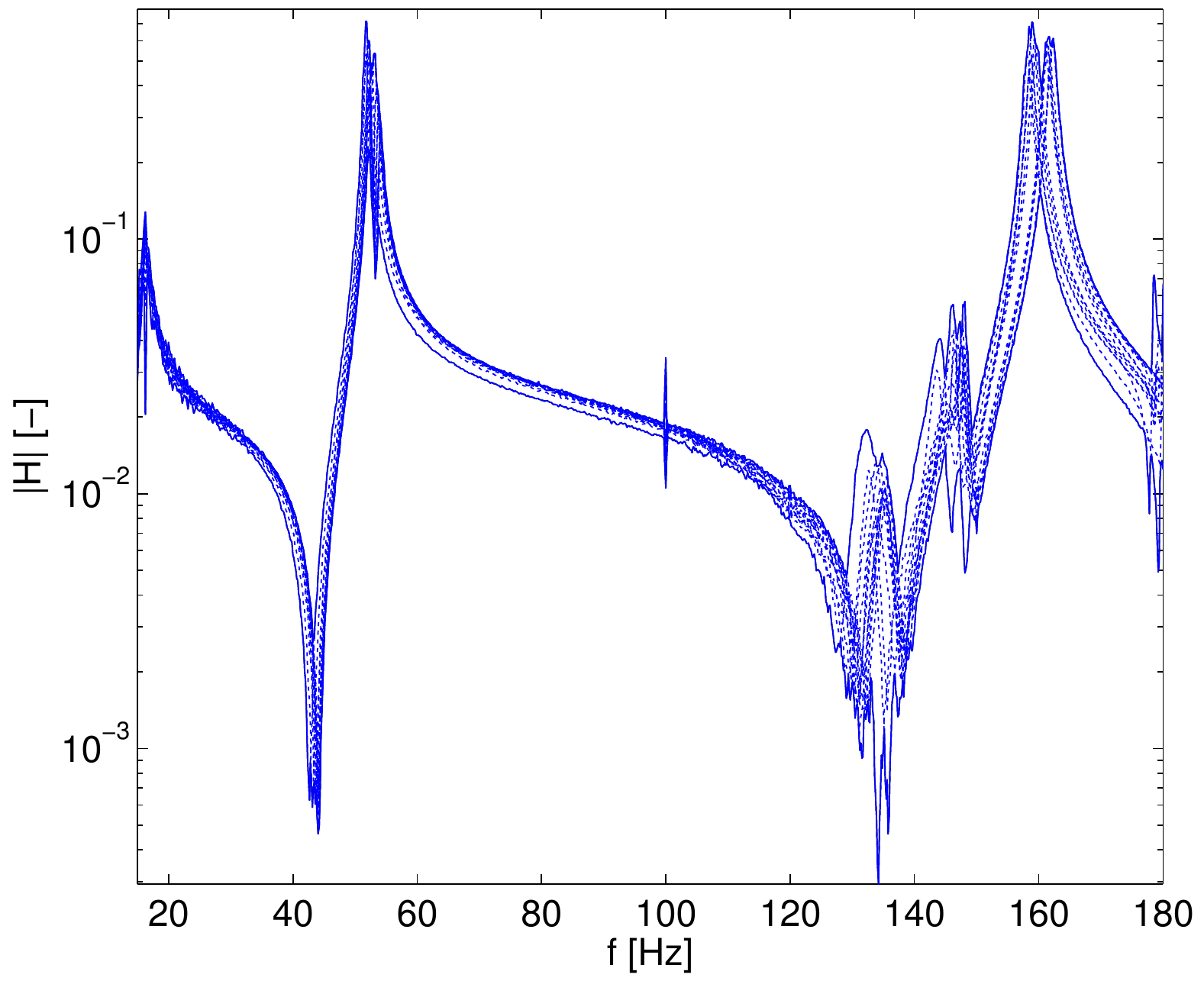}
	\caption{Measured FRFs. Dotted lines mark the individual measurements.}
	\label{fig:FRF_measured_all}
\end{figure}
\begin{figure}
	\centering
		\includegraphics[width=0.5\textwidth]{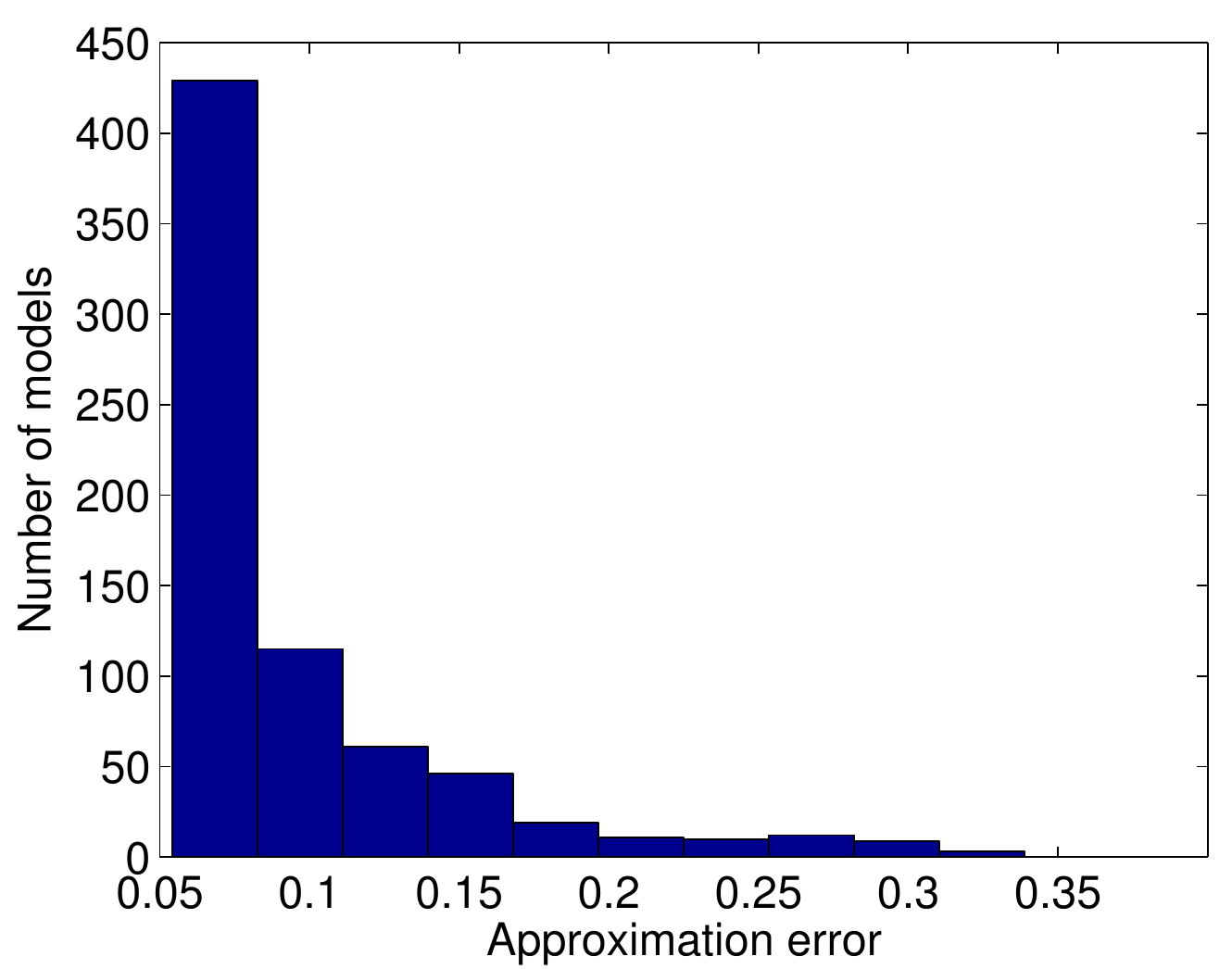}
  \caption{Histogram of the approximation errors}
	\label{fig:histogram_error}
\end{figure}

To introduce an uncertainty in the measurements, an additional mass of $0.9 kg$ is mounted to the truss structure at different positions. The additional mass corresponds to approximately 1.5\% of the mass of the truss structure (60 kg).
In sum, 14 different configurations are considered: First without the steel mass, and further on with the mass mounted to the 13 possible positions on top of the truss, excepting the position of the actuator. The acquired frequency response functions are plotted in Figure \ref{fig:FRF_measured_all}. The added mass decreased the first resonance frequency at $16.1 Hz$ by maximal 14\% and the resonance frequency at $162.3 Hz$ by 2.3\%.
One of the measurements (FRF number 12) is used for validation, i.e. as the test case. Four measurements are used as training cases to find the poles for the Kautz filter bank. By permutation 715 different configurations of four training measurements can be defined. The procedure described above in figure \ref{kautz_procedure} was applied to all the sets.\\
To this end, the four impulse responses are used in a modified Prony algorithm as mentioned above, which treats the data set as a multiple input single output system \cite{williamson_globally_1996}. The analysis is repeated for models of increasing order, delivering basis sets of Kautz poles. For higher orders, however, the Prony algorithm procures poles far away from the real resonance frequencies, which are not useful as parameters of the Kautz filters. Thus, the set of poles gathered from a Prony analysis of order six, which matches the number of peaks in the FRF, was periodically repeated in order to realize a higher order parameter set.
The approximation error obviously depends to the considered training set. In the best case, an error of 0.05 can be gained at the highest considered model order of 80, while for the worst case an error of 0.34 is observed (Figure \ref{app_error_exp}, left). The mean value, however, is well below the worst case, and the histogram of the  approximation error for all 715 cases indicates, that for most cases good models with an error below 0.1 can be gained (Figure \ref{fig:histogram_error}). The identification of the test case is performed again with FIR filters of ascending order (Figure \ref{app_error_exp}, left). The approximation errors are compared with those obtained with the Kautz filters. For an error of 0.05 as gained from the best experiment, an FIR filter of order 800 is required, which would mean a significantly higher effort for the implementation compared to the Kautz filter of order 80. To obtain an error of 0.34, as observed in the worst case with the 80th order Kautz filter, an FIR filter of order 250 is needed, meaning that for this case a similar computation effort is needed for both filters. 
%
%
\begin{figure}
\begin{tabular}{cc}
\includegraphics[width=0.5\textwidth]{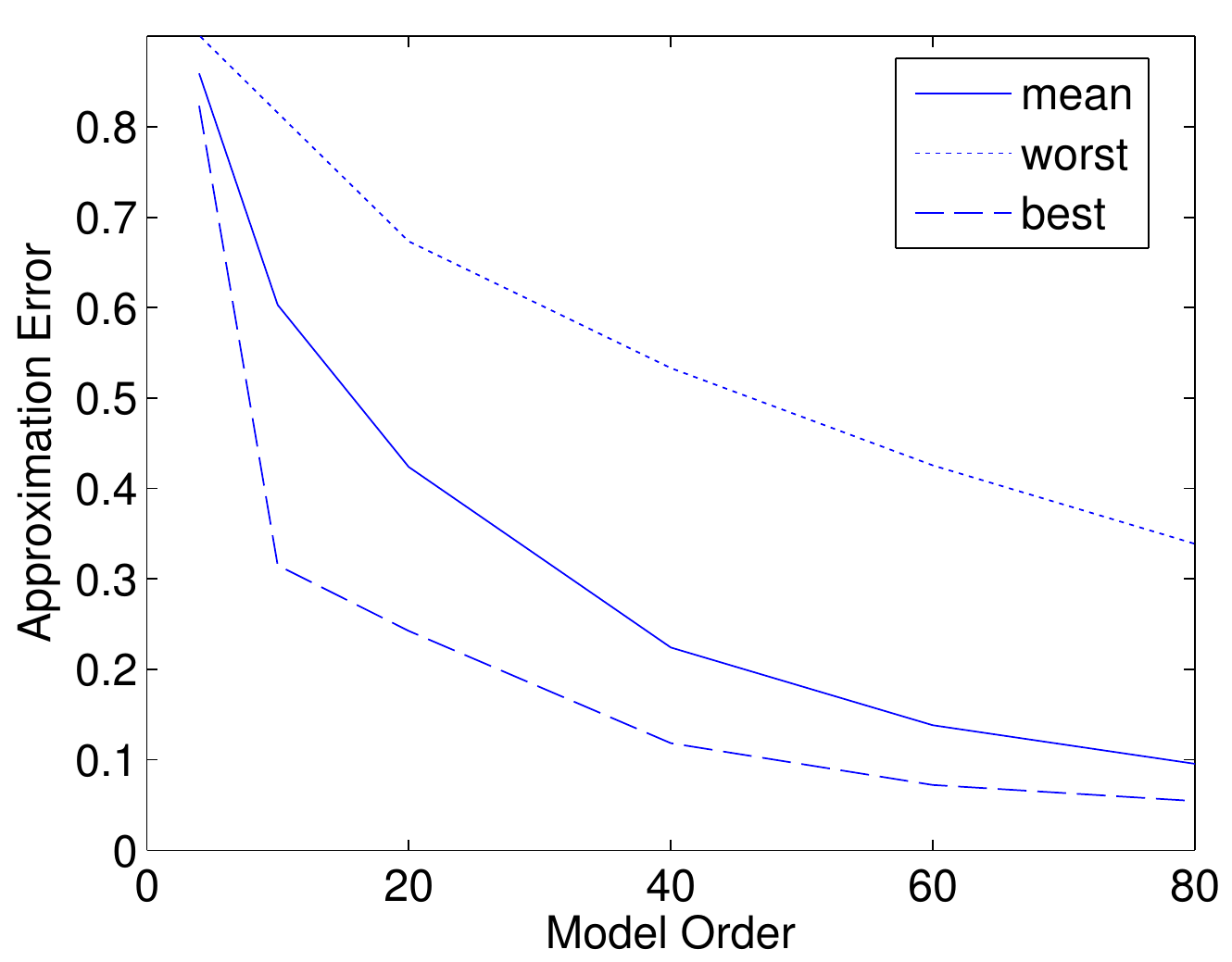}&\includegraphics[width=0.5\textwidth]{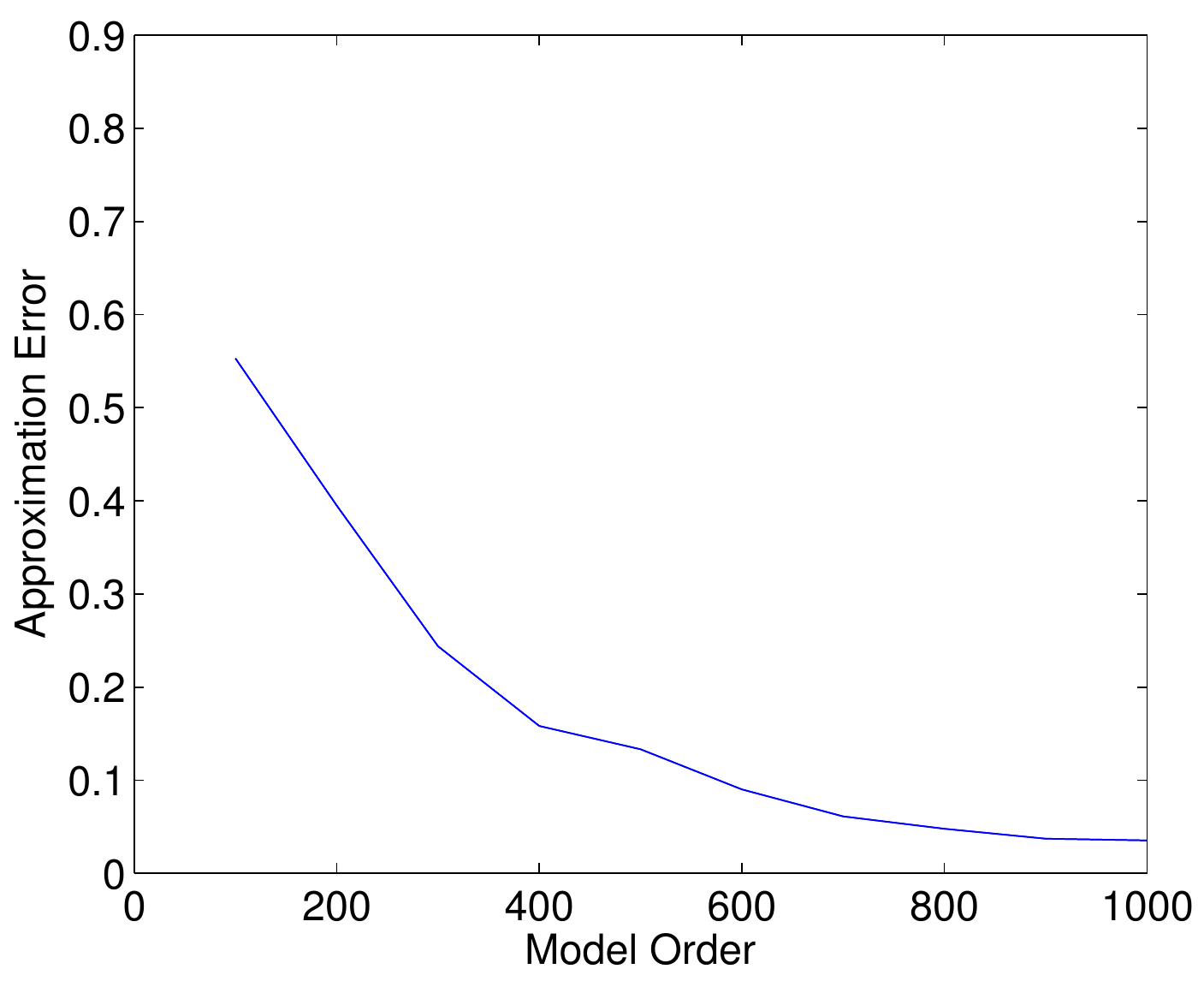}
\end{tabular}
\caption{Approximation of a truss structure FRF with Kautz filters of ascending orders (left) and FIR filters (right)}
\label{app_error_exp}
\end{figure}

\begin{figure}
\begin{tabular}{cc}
\includegraphics[width=0.5\textwidth]{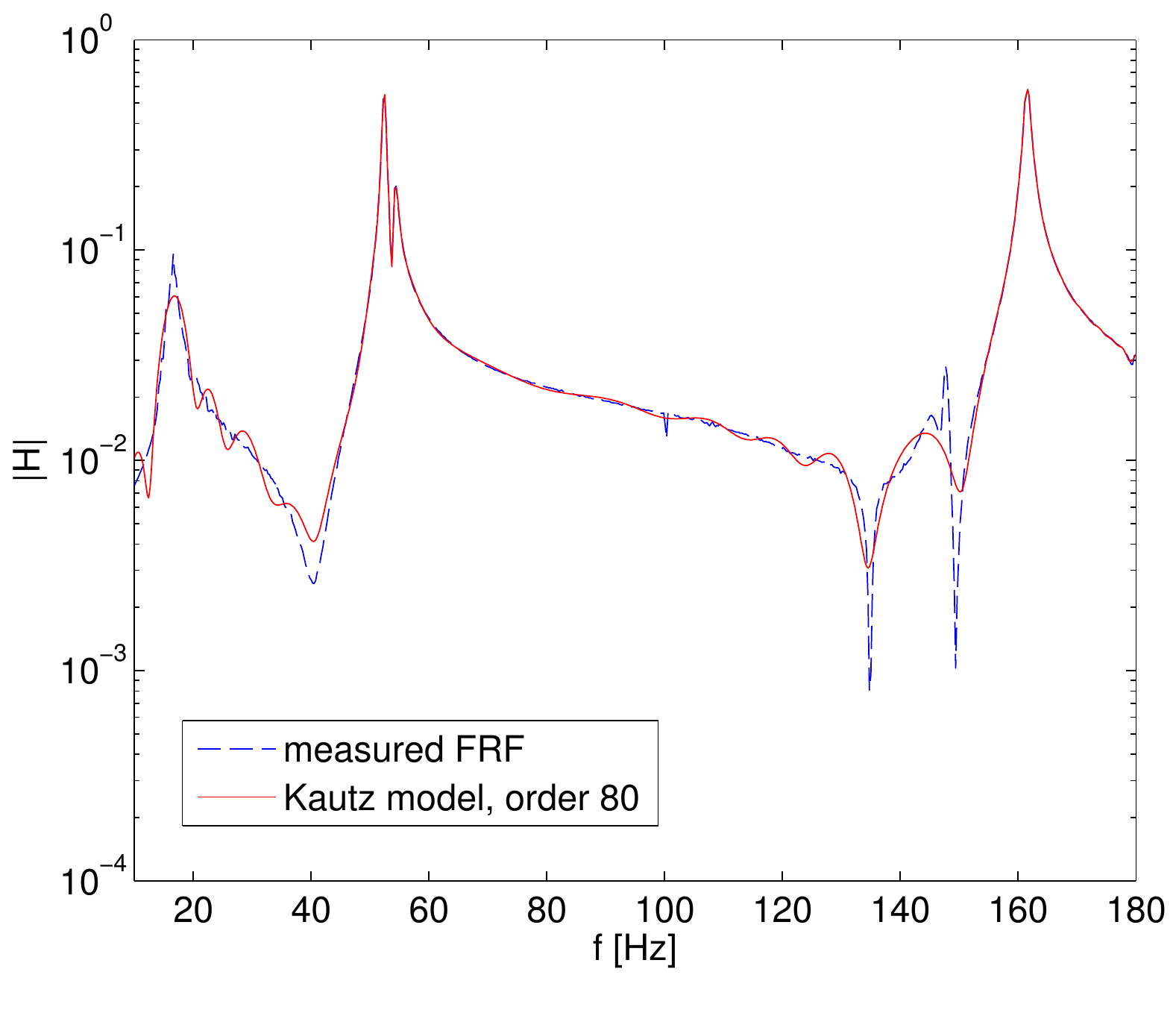}&\includegraphics[width=0.5\textwidth]{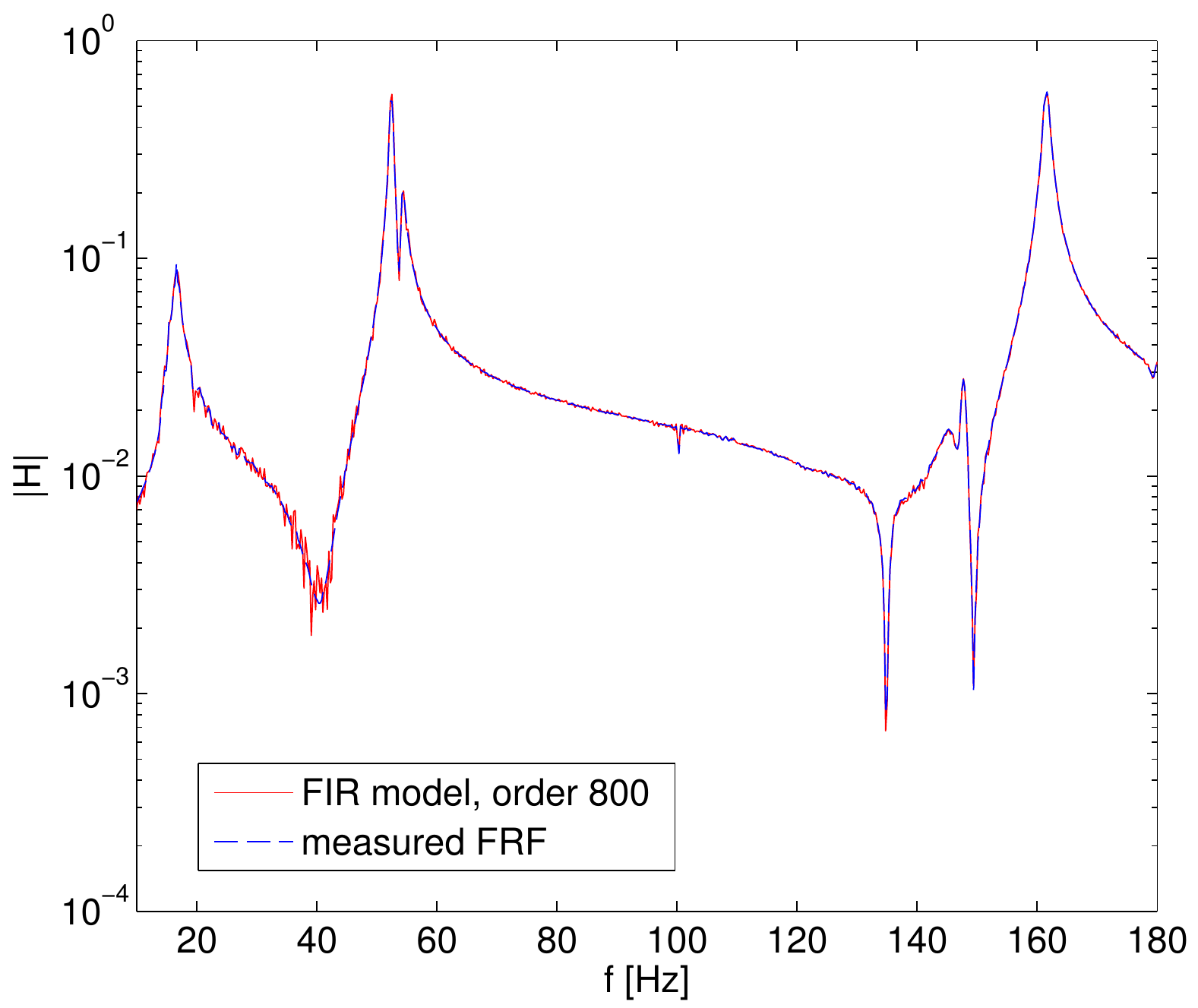}
\end{tabular}
\caption{Approximation of a truss structure FRF with Kautz filters of ascending orders for the best case (left) and FIR filters (right)}
\label{app_error_exp}
\end{figure}


\section{Summary and Conclusions}
In this work a study on the system identification with orthonormal filters was presented. The goal of the study was to examine the performance of this class of filters when the necessary a priori knowledge contains a certain level of uncertainty, i.e. the estimated poles deviate from those of the actual system.\\
An investigation on a single-degree-of-freedom oscillator, which can be considered the basic problem for vibrating structures, indicated that the Kautz filter bank can represent this system with a significantly lower effort than an FIR filter, if a number of example systems is available to gather the a priori knowledge. The model error was deteriorated when the poles of the example systems were deviating from the poles of the test system to be identified. These results could be validated by experiments conducted at a laboratory set up. Since the experimental set up was a multiple-degree-of-freedom system with several resonances, the saving of computational burden was less significant than in the case of the SDOF system. Thus, the orthonormal filter banks can especially be supposed to be an efficient alternative to FIR filters when a low number of those resonances has to be considered.\\
Further work can be conducted on pole selection strategies for the filter banks from the identification of the example systems with respect to the performance and required number of filter bank stages. A very relevant topic is also the influence of the variability of the system parameters on the needed filter order. It can be supposed, that an adaptive Kautz filter will have more coefficients, if the variability of the systems rises. This should be subject of further investigations.
\begin{acknowledgement}
The work was partly supported by the LOEWE centre AdRIA funded by the state of Hesse. The support is gratefully acknowledged.
\end{acknowledgement}

\bibliographystyle{unsrt}
\bibliography{Biblio}

\end{document}